\documentclass{ws-procs9x6}
\newcommand{\be}{\begin{equation}}
\newcommand{\ee}{\end{equation}}
\newcommand{\bea}{\begin{eqnarray}}
\newcommand{\eea}{\end{eqnarray}}

\begin{document}
%\rightline{BARI-TH/541-06}
\title{CONSTRAINING UNIVERSAL EXTRA DIMENSIONS\\ THROUGH B DECAYS}

\author{F. DE FAZIO$^*$}

\address{Istituto Nazionale di Fisica Nucleare, Sezione di Bari,
\\Via Orabona 4, I-70126 Bari Italy\\
$^*$E-mail: fulvia.defazio@ba.infn.it}

\begin{abstract}
We analyze the exclusive  rare  $B \to K^{(*)} \ell^+ \ell^-$, $B
\to K^{(*)} \nu \bar \nu$ and $B \to K^* \gamma$ decays in the
Applequist-Cheng-Dobrescu model,  an extension of the Standard
Model in presence of universal extra dimensions. In  the case of a
single universal extra dimension, we study the dependence of
several observables on the compactification parameter $1/R$, and
discuss  whether the hadronic uncertainty due to the  form factors
obscures or not such a dependence. We find that, using present
data, it is possible in many cases to put a sensible lower bound
to  $1/R$, the most stringent one coming from $B \to K^* \gamma$.
\end{abstract}


\bodymatter

\section{Introduction} \label{sec:intro}
Rare B decays induced by $b \to s$ transition play a peculiar role
in searching for new Physics, being  induced at loop level and
hence  suppressed in the Standard Model (SM)\cite{Hurth:2003vb}.
They can also be useful to constrain  extra dimensional
scenarios\cite{Agashe:2001xt}. This is  the case of
  the Appelquist-Cheng-Dobrescu (ACD)
  model\cite{Appelquist:2000nn} in which universal extra dimensions are
considered, which means that
 all the fields are allowed to
 propagate in all available dimensions.
  In the case of  a single  extra
dimension compactified on a circle of radius R, Tevatron run I
data  allow to put the bound $1/R \ge 300 $ GeV. To be more
general, we analyze a broader range $1/R \geq 200$ GeV.

In Refs. \refcite{Buras:2002ej,Buras:2003mk} the effective
hamiltonian relative to   inclusive $b \to s$ decays was computed
within the ACD model. In this paper, we summarize the results
obtained in Ref. \refcite{Colangelo:2006vm} for exclusive $b \to
s$-induced modes.
 In this case, the uncertainty in the  form factors must be considered,
  since it can  overshadow the sensitivity to the compactification
 parameter $1/R$. Indeed we find that
 computing the  branching
 ratios of $B \to K^{(*)} \ell^+  \ell^-$ and the forward-backward lepton asymmetry
 in $B \to K^* \ell^+  \ell^-$ for a representative set of form factors,  a bound can
 be put.
We  also study the modes $B \to K^{(*)} \nu \bar \nu$, for which
no signal has been observed, so far, and the
  $BR(B \to K^* \gamma)$ versus  $1/R$,  which allows  to
 establish the most stringent  bound on $1/R$.

\section{The ACD model with a single UED} \label{sec:acd}
The  ACD model\cite{Appelquist:2000nn}   consists in the minimal
extension of the SM in $4+ \delta$ dimensions; we  consider
 $\delta=1$. The fifth dimension $x_5=y$ is compactified
to the orbifold $S^1/Z_2$, i.e. on a circle of radius R
  and runs
from 0 to $ 2 \pi R$ with   $y=0, y=\pi R$  fixed points of the
orbifold. Hence a field $F(x,y)$ ($x$ denoting the  usual 3+1
coordinates) would be a
 periodic function of $y$,  and  it could be expressed as
 $\displaystyle{F(x,y)=\sum_{n=- \infty}^{n=+  \infty} F_n(x) e^{i\, n \cdot
 y/R}}$.
If $F$ is  a massless boson field, the KK modes $F_n$  obey the
equation
 $\displaystyle{\left(\partial^\mu
\partial_\mu+ n^2/ R^2 \right)F_n(x)=0 }$,
 $\mu=0,1,2,3$ so that, apart the zero mode, they
 behave in four dimensions as massive  particles with $m_n^2= (n/R)^2$.
Under the parity transformation $P_5: y \to -y$ fields having a
correspondent in
 the 4-d SM should be even, so that their zero mode in the  expansion is interpreted
as the ordinary SM field. On the other hand, fields having no
 SM partner should be odd,  so that  they do not
have zero modes.

Important features of the ACD model are: i) there is
 a single  additional free parameter with respect to the SM,
the compactification radius $R$; ii)  conservation of KK parity,
with the consequence that there is no tree-level contribution of
KK modes in low energy processes (at scales
 $\mu \ll 1/R$) and no production of single KK excitation in ordinary particle
 interactions.
A detailed description  of  this model is provided in Ref.
\refcite{Buras:2002ej}.

\section{Decays $ B \to K^{(*)} \ell^+ \ell^-$ } \label{sec:modes}

In the Standard Model the effective $ \Delta B =-1$, $\Delta S =
1$ Hamiltonian governing  the  transition $b \to s \ell^+ \ell^-$
is $ H_W\,=\,4\,{G_F \over \sqrt{2}} V_{tb} V_{ts}^*
\sum_{i=1}^{10} C_i(\mu) O_i(\mu)$. \noindent $G_F$ is the Fermi
constant and $V_{ij}$ are elements of the
Cabibbo-Kobayashi-Maskawa mixing matrix; we neglect terms
proportional to $V_{ub} V_{us}^*$. $O_1$, $O_2$ are
current-current operators, $O_3,...,O_6$   QCD penguins, $O_7$,
$O_8$  magnetic penguins, $O_9$, $O_{10}$ semileptonic electroweak
penguins. We do not consider
 the contribution to $B \to K^{(*)} \ell^+ \ell^-$  with the  lepton pair  coming from
 $c{\bar c}$ resonances,  mainly due to  $O_1$, $O_2$.
We also neglect QCD penguins whose  coefficients are very small
compared to the others. Therefore, in the case of the modes $B \to
K^{(*)} \ell^+ \ell^-$, the relevant operators are: $O_7={e \over
16 \pi^2} m_b ({\bar s}_{L \alpha} \sigma^{\mu \nu}
     b_{R \alpha}) F_{\mu \nu} $,
$O_9={e^2 \over 16 \pi^2}  ({\bar s}_{L \alpha} \gamma^\mu
     b_{L \alpha}) \; {\bar \ell} \gamma_\mu \ell $, $
O_{10}={e^2 \over 16 \pi^2}  ({\bar s}_{L \alpha} \gamma^\mu
     b_{L \alpha}) \; {\bar \ell} \gamma_\mu \gamma_5 \ell
$. Their coefficients   have been computed at NNLO in the Standard
Model\cite{nnlo} and at
 NLO   for  the ACD model\cite{Buras:2002ej,Buras:2003mk}: we
 use these results in our study. No new operators  are found in
 ACD, while
  the   coefficients  are modified because particles not
present in  SM can contribute as intermediate states in loop
diagrams. As a consequence, they are expressed in terms of
functions $F(x_t,1/R)$, $x_t=m_t^2/M_W^2$,  generalizing the
corresponding SM functions $F_0(x_t)$ according to
$\displaystyle{F(x_t,1/R)=F_0(x_t)+\sum_{n=1}^{\infty}F_n(x_t,x_n)}$,
 where $x_n=m_n^2/M_W^2$
and $m_n=n/R$\cite{Buras:2002ej,Buras:2003mk}.
 For large values of $1/R$ the SM phenomenology should be
  recovered, since the new states,  being more and more massive,  decouple from the low-energy theory.

The exclusive $B \to K^{(*)} \ell^+ \ell^-$ modes involve
  the matrix elements of the operators  in
the effective hamiltonian  between the $B$ and  $K$ or $K^*$
mesons, for which we use the standard parametrization in terms of
form factors: \bea <K(p^\prime)|{\bar s} \gamma_\mu b
|B(p)>=(p+p^\prime)_\mu F_1(q^2) +{M_B^2-M_K^2 \over q^2} q_\mu
\left
(F_0(q^2)-F_1(q^2)\right )\hskip 3 pt ; \nonumber \\
<K(p^\prime)|{\bar s}\; i\;  \sigma_{\mu \nu} q^\nu b |B(p)>=
\Big[(p+p^\prime)_\mu q^2 -(M_B^2-M_K^2)q_\mu\Big] \; {F_T(q^2)
\over M_B+M_K} \hskip 3 pt ; \nonumber  \eea
\begin{eqnarray}
&&<K^*(p^\prime,\epsilon)|{\bar s} \gamma_\mu (1-\gamma_5) b
|B(p)>= \epsilon_{\mu \nu \alpha \beta} \epsilon^{* \nu} p^\alpha
p^{\prime \beta}
{ 2 V(q^2) \over M_B + M_{K^*}}  \nonumber \\
&-& i \left [ \epsilon^*_\mu (M_B + M_{K^*}) A_1(q^2) -
(\epsilon^* \cdot q) (p+p')_\mu  {A_2(q^2) \over (M_B + M_{K^*}) }
\right. \nonumber \\ &-& \left. (\epsilon^* \cdot q) {2 M_{K^*}
\over q^2} \big(A_3(q^2) - A_0(q^2)\big) q_\mu \right ] \hskip 3
pt ;\nonumber
\end{eqnarray}
\begin{eqnarray}
&&<K^*(p^\prime,\epsilon)|{\bar s} \sigma_{\mu \nu} q^\nu
{(1+\gamma_5) \over 2} b |B(p)>= i \epsilon_{\mu \nu \alpha \beta}
\epsilon^{* \nu} p^\alpha p^{\prime \beta}
\; 2 \; T_1(q^2)  + \nonumber \\
&+&  \Big[ \epsilon^*_\mu (M_B^2 - M^2_{K^*})  -
(\epsilon^* \cdot q) (p+p')_\mu \Big] \; T_2(q^2) \nonumber \\
&+& (\epsilon^* \cdot q) \left [ q_\mu - {q^2 \over M_B^2 -
M^2_{K^*}} (p + p')_\mu \right ] \; T_3(q^2) \hskip 3 pt
,\nonumber
\end{eqnarray}
\noindent  where $q=p-p^\prime$, $\displaystyle{ A_3(q^2) = {M_B +
M_{K^*} \over 2 M_{K^*}} A_1(q^2) - {M_B - M_{K^*} \over 2
M_{K^*}} A_2(q^2)}$ with the conditions $F_1(0)=F_2(0)$, $ A_3(0)
= A_0(0)$, $T_1(0) = T_2(0)$.

 We use two sets of  form factors:
 the first one (set A) obtained by  three-point QCD sum rules
based on the short-distance expansion\cite{Colangelo:1995jv};  the
second one (set B) obtained by light-cone QCD sum
rules\cite{Ball:2004rg}.  For both sets we include in the
numerical analysis the errors on the parameters.

\begin{figure}[ht]
\begin{center}\vskip -0.3cm
\includegraphics[width=0.45\textwidth] {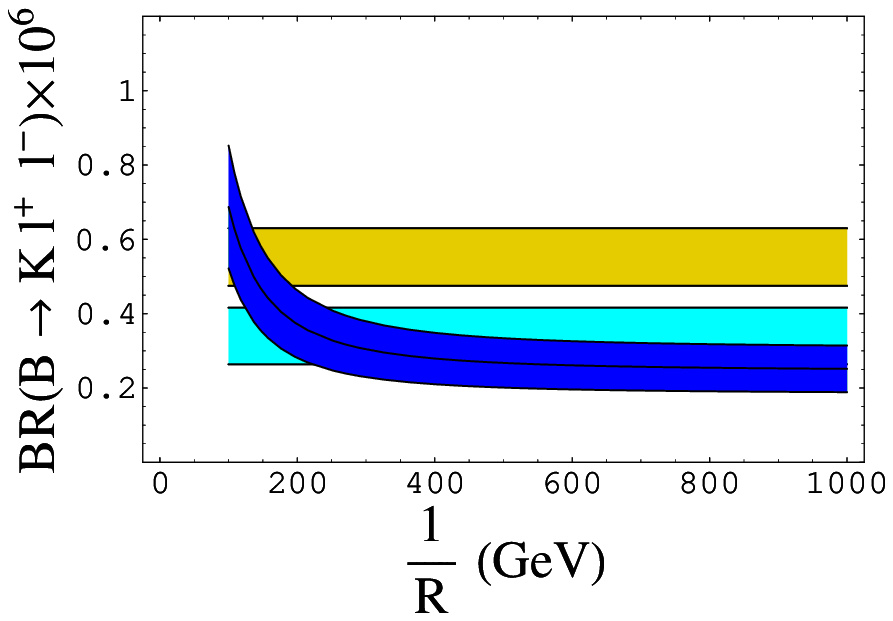} \hspace{0.5cm}
 \includegraphics[width=0.45\textwidth] {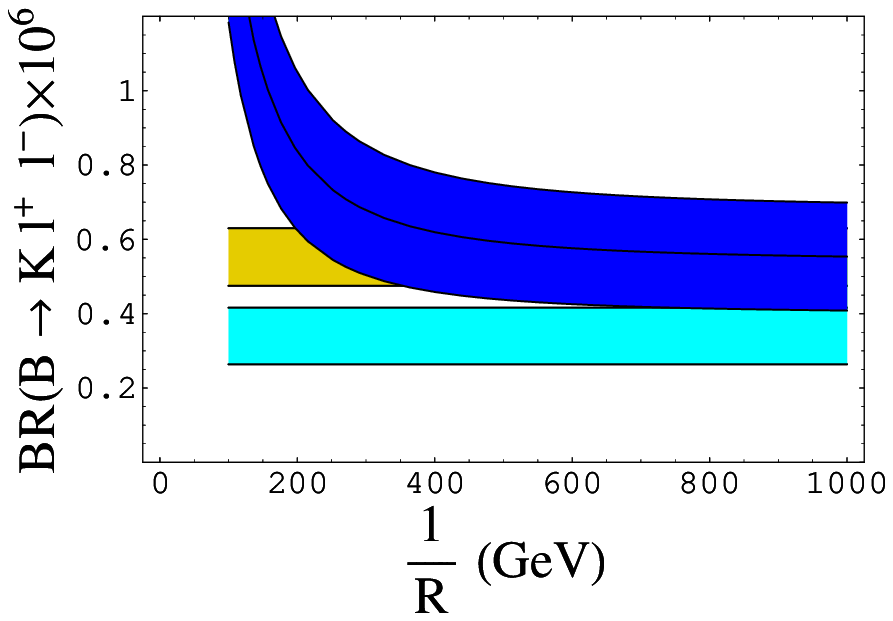}\\
 \includegraphics[width=0.45\textwidth] {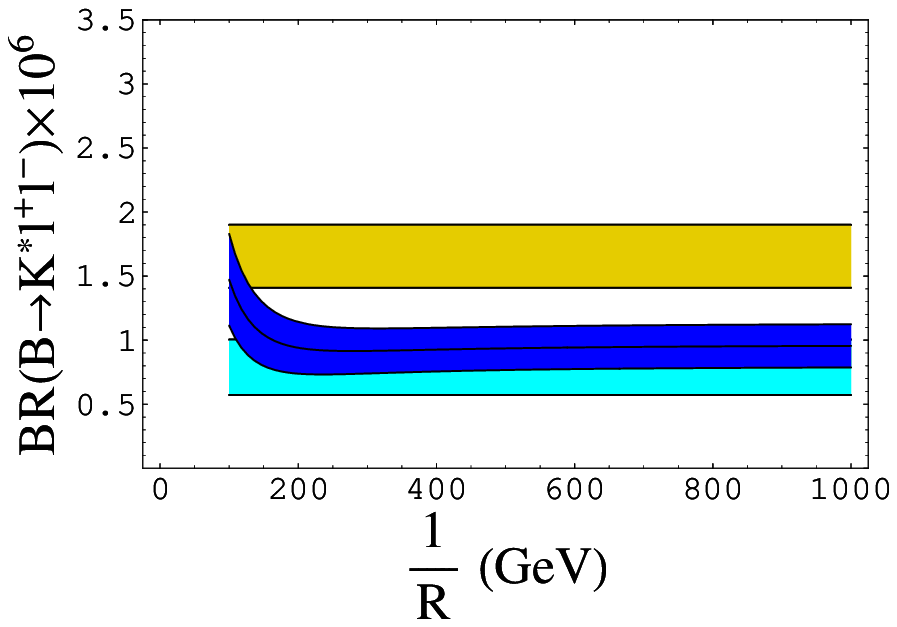} \hspace{0.5cm}
\includegraphics[width=0.45\textwidth] {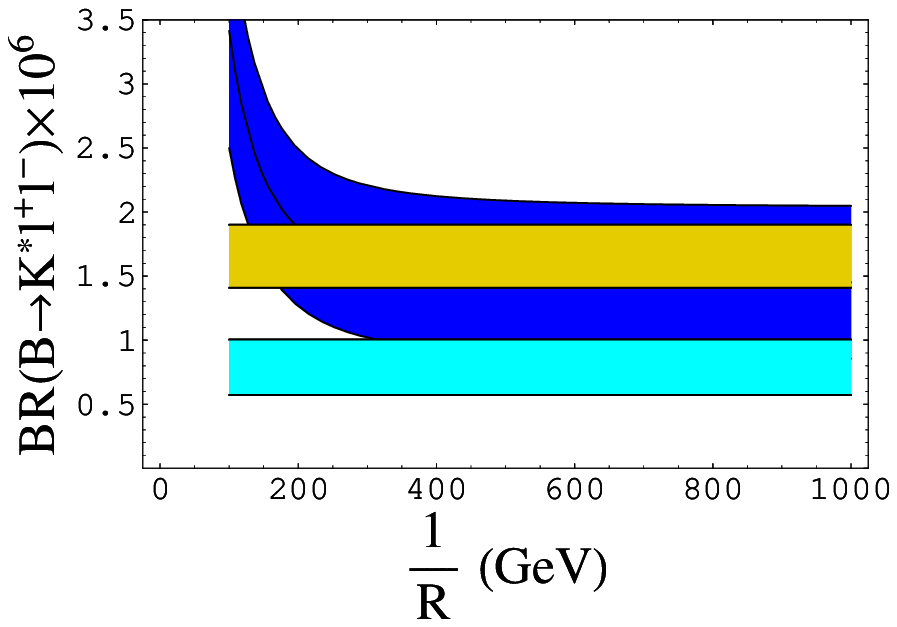}
\end{center}
\caption{\baselineskip=15pt  $BR(B \to K \ell^+ \ell^-)$ (upper
figures) and $BR(B \to K^* \ell^+ \ell^-)$ (lower figures)  versus
$1/R$ using set A (left) and B
 (right) of form factors.  The two horizontal regions refer
to  BaBar (lower band) and Belle (upper band) data.}
\vspace*{1.0cm} \label{brkcol}
\end{figure}

 In Fig. \ref{brkcol} we
plot, for the two sets of form factors,  the branching fractions
relative to $B \to K^{(*)} \ell^+ \ell^-$ versus $1/R$
 and compare them with the experimental data provided by
BaBar\cite{Aubert:2004it,Aubert:2005cf} and
Belle\cite{Iwasaki:2005sy,Abe:2004ir}: \bea BR(B\to K \ell^+
\ell^- )&=&(5.50 \pm^{0.75}_{0.70} \pm 0.27 \pm 0.02) \times
10^{-7}\,\,\, (Belle)  \nonumber \\ &=&(3.4 \pm 0.7 \pm 0.3)
\times 10^{-7} \,\,\,(BaBar) \nonumber \\  BR(B \to K^* \ell^+
\ell^-) &=&  (16.5 \pm^{2.3}_{2.2} \pm 0.9 \pm 0.4)\times
10^{-7}\,\,\,(Belle) \nonumber \\ &=& (7.8 \pm^{1.9}_{1.7} \pm
1.2)\times
      10^{-7} \,\,\,(BaBar).  \eea
Set B
 excludes  $1/R \leq 200$ GeV.
  Improved
data will resolve the  discrepancy between the  experiments and
increase the lower bound for $1/R$.

\begin{figure}[ht]
\begin{center}\vskip -0.7cm
\includegraphics[width=0.45\textwidth] {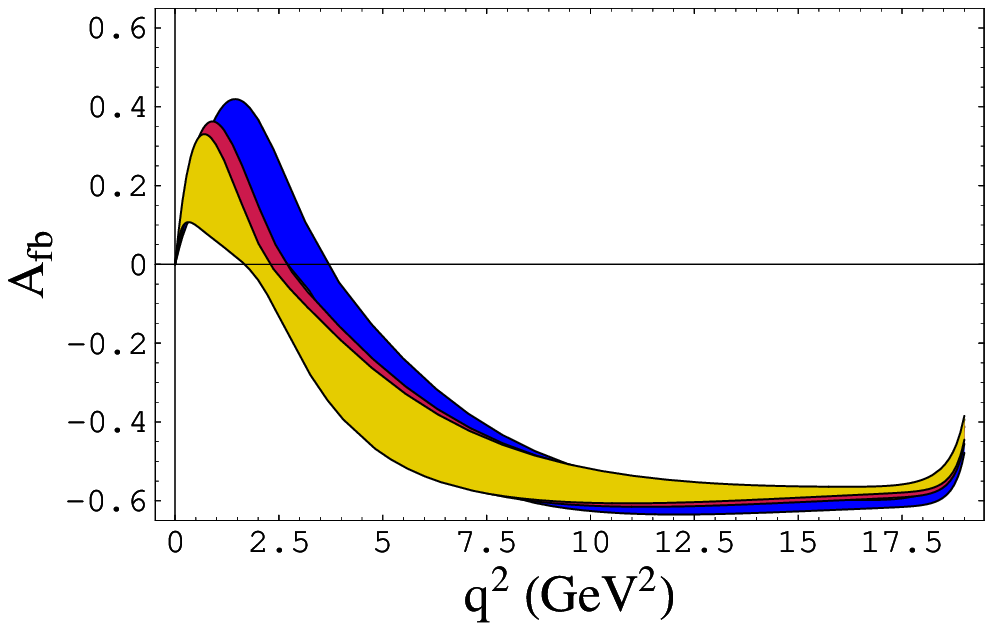} \hspace{0.5cm}
\includegraphics[width=0.45\textwidth] {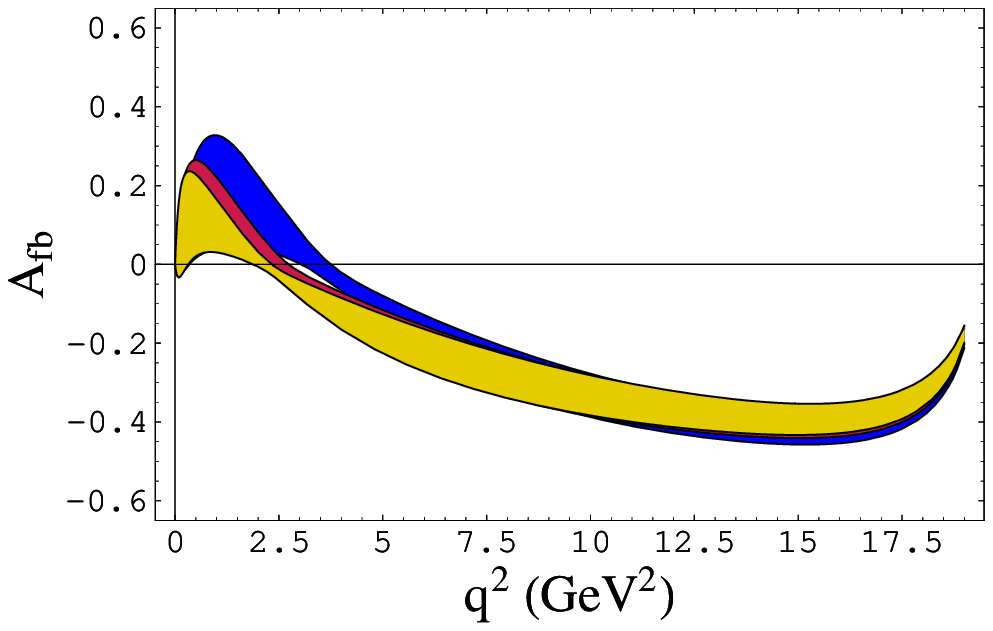}
\end{center}
\caption{\baselineskip=15pt Forward-backward lepton asymmetry in
$B \to K^* \ell^+ \ell^-$  versus $1/R$ using set A  (left) and B
(right). The  dark (blue) bands correspond to the SM results, the
intermediate (red) band to $1/R=250$ GeV, the light (yellow) one
to
 $1/R=200$ GeV. } \vspace*{1.0cm} \label{fig-afb}
\end{figure}

In the case of  $ B \to K^* \ell^+ \ell^-$ the  investigation of
the forward-backward asymmetry ${\cal A}_{fb}$ in the dilepton
angular distribution may also reveal effects beyond the SM. In
particular,
 in  SM,  due to the opposite sign of the
coefficients $C_7$ and $C_9$, ${\cal A}_{fb}$ has a zero the
position of which is almost independent of the model for  the form
factors\cite{Burdman:1998mk}. Let  $\theta_\ell$ be the angle
between the $\ell^+$ direction and the $B$ direction in the rest
frame of the lepton pair (we consider  massless leptons). We
define:
\begin{equation}
{\cal A}_{fb} (q^2)=\displaystyle {\displaystyle \int_0^1{d^2
\Gamma \over dq^2 d cos\theta_\ell}dcos\theta_\ell
-\int^0_{-1}{d^2 \Gamma \over dq^2 d
cos\theta_\ell}dcos\theta_\ell \over\displaystyle \int_0^1{d^2
\Gamma \over dq^2 d cos\theta_\ell}dcos\theta_\ell
+\int^0_{-1}{d^2 \Gamma \over dq^2 d
cos\theta_\ell}dcos\theta_\ell} \; .
 \label{asim_def}
\end{equation}
 We
show in Fig. \ref{fig-afb} the
 predictions for the SM,  $1/R=250$ GeV  and
$ 1/R=200$ GeV. The zero of ${\cal A}_{fb}$ is  sensitive
 to the compactification parameter, so that  its experimental determination  would  constrain $1/R$.
At present, the analysis performed by Belle Collaboration
indicates that the relative sign of  $C_9$ and $C_7$ is negative,
confirming that ${\cal A}_{fb}$ should have a
zero\cite{Abe:2005km}.
%%%%%%%%%%%%%%%%%%%%%%%%%%%%%%%%%%%%%%%%%%
%*****************
% ************************************************************************ ***
%*****************
%

\section{The decays $B \to K^{(*)} \nu {\bar \nu}$}
%************************************************************************ ***

 In the
SM the effective Hamiltonian  governing    $ b \to s \nu {\bar
\nu} $  induced decays is  \be {\cal H}_{eff} = \displaystyle{G_F
\over \sqrt 2} {\alpha \over 2 \pi \sin^2(\theta_W)} \; V_{ts}
V^*_{tb} \; \eta_X X(x_t) \; {\bar b} \gamma^\mu ( 1- \gamma_5) s
\; {\bar \nu} \gamma_\mu ( 1- \gamma_5) \nu \label{hamilnunubar}
\ee obtained from $Z^0$ penguin and box diagrams dominated by
 the intermediate top quark. In
(\ref{hamilnunubar})  $\theta_W$ is the Weinberg angle. We put to
unity the QCD factor
$\eta_X$\cite{buchalla,buchalla1,Buchalla:1998ba}.The function $X$
was computed in Refs. \refcite{inami,buchalla,buchalla1} in the SM
and in the ACD model in Refs. \refcite{Buras:2002ej,Buras:2003mk}.

$B \to K^{(*)} \nu {\bar \nu}$ decays have been studied within the
SM\cite{Colangelo:1996ay,Buchalla:2000sk}, while  in Ref.
\refcite{Colangelo:2006vm} the dependence of $BR(B \to K \nu \bar
\nu )$ and $BR(B \to K^* \nu \bar \nu )$ on $1/R$ has been
derived. However,
 only an experimental upper bound exists  for  $B \to K
\nu \bar \nu$: $BR(B^- \to K^- \nu \bar \nu)  < 3.6 \times 10^{-5}
\,\,(90 \% \,\, CL)$ \cite{Abe:2005bq},  $BR(B^- \to K^- \nu \bar
\nu)<5.2 \times 10^{-5} \,\,\,\,(90 \% \,\,
CL)$\cite{Aubert:2004ws}, furthermore the $1/R$ dependence turns
out to be  too mild for distinguishing values above $1/R \ge 200$
GeV.

%*****************
\section{The decay $ B \to K^* \gamma$ }
\begin{figure}[ht]
\begin{center}
\includegraphics[width=0.45\textwidth] {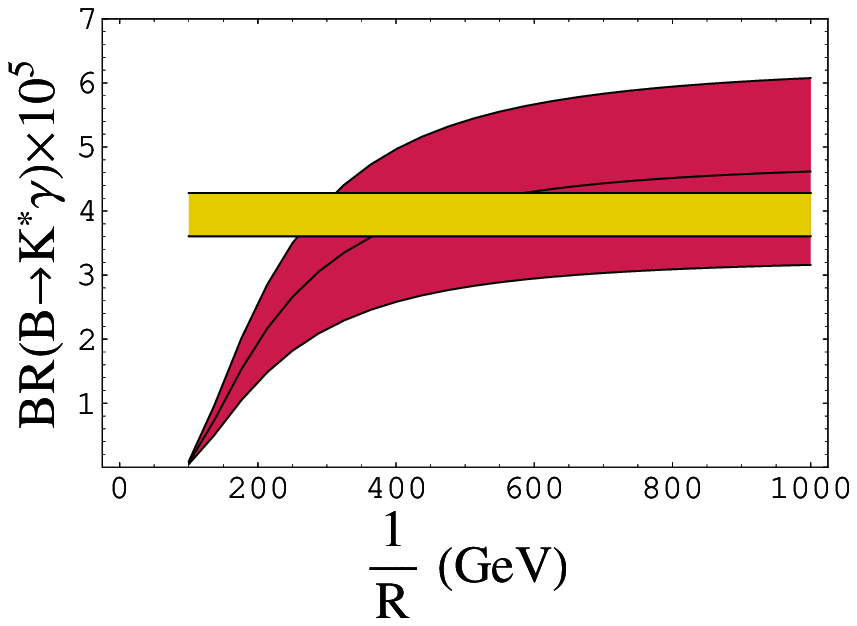} \hspace{0.5cm}
 \includegraphics[width=0.45\textwidth] {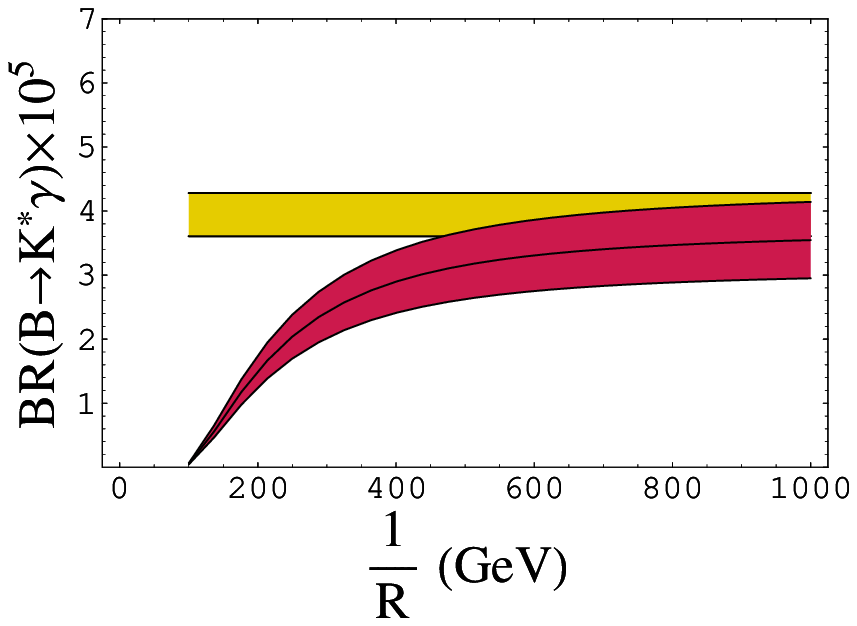}
\end{center}
\caption{\baselineskip=15pt $BR(B \to K^* \gamma)$  versus $1/R$
using  set A  (left) and   B
 (right) of form factors .  The  horizontal
band corresponds to the experimental result. } \vspace*{1.0cm}
\label{brkstargamma}
\end{figure}

The transition $b \to s \gamma$ is described  by the operator
$O_7$. The most recent measurements for the exclusive branching
fractions are\cite{Nakao:2004th,Aubert:2004te}:\bea
 BR(B^0 \to K^{*0} \gamma)&=&(4.01 \pm
0.21 \pm 0.17 ) \times 10^{-5} \,\,\,(Belle)\nonumber \\ &=&(3.92
\pm 0.20\pm 0.24 )
 \times 10^{-5}    \,\,\, (BaBar) \nonumber \\ BR(B^-
\to K^{*-} \gamma)&=&(4.25 \pm 0.31 \pm 0.24) \times 10^{-5}
\,\,\,
 (Belle)\nonumber \\
   &=&(3.87 \pm 0.28 \pm 0.26 ) \times 10^{-5}\,\,\, (BaBar)
 \nonumber \eea In Fig. \ref{brkstargamma} the
 branching ratio computed in the ACD model
 is plotted versus  $1/R$: the sensitivity to the
 this
 parameter is evident;  a lower bound of
  $1/R \ge 250$ GeV  can be put adopting set A, and a stronger bound of    $1/R \ge
400$ GeV using set B, which is the most stringent bound that can
be currently put on this parameter from the  $B$ decay modes we
have considered.
%%%%%%%%%%%%%%%%%%%%%%%%%%%%%%%%%%%%%%%%%%
%*****************

\section{Conclusions and Perspectives} \label{sec:concl}
We have shown how  the predictions for    $B \to K^{(*)} \ell^+
\ell^-$, $B \to K^{(*)}\nu \bar \nu$, $B \to K^* \gamma$ decays
are modified within the ACD scenario. The constraints on  $1/R$
are
  slightly model dependent,   being
different using different sets of form factors. Nevertheless,
 various distributions, together with the lepton
forward-backward asymmetry in $B \to K^* \ell^+ \ell^-$ are very
promising in order to constrain $1/R$,  the most stringent lower
bound coming from $B \to K^* \gamma$. Improvements in the
experimental data, expected in the near future, will allow to
establish more stringent constraints for the compactification
radius.

\section*{Acknowledgments}
I warmly thank the organizers  of the workshop for their kind
hospitality.  I am grateful to P. Colangelo, R. Ferrandes and T.N.
Pham for collaboration on the analyses discussed above and I
acknowledge partial support from the EC Contract No.
HPRN-CT-2002-00311 (EURIDICE).


\begin{thebibliography}{9}
%\cite{Hurth:2003vb}
\bibitem{Hurth:2003vb}
 For a  review see T.~Hurth,
  %``Present status of inclusive rare B decays,''
  Rev.\ Mod.\ Phys.\  {\bf 75}, 1159 (2003) and Refs. therein.
 % [arXiv:hep-ph/0212304].
  %%CITATION = HEP-PH 0212304;%%


%\cite{Agashe:2001xt}
\bibitem{Agashe:2001xt}
  K.~Agashe, N.~G.~Deshpande and G.~H.~Wu,
  %``Universal extra dimensions and b $\to$ s gamma,''
  Phys.\ Lett.\ B {\bf 514}, 309 (2001).
%  [arXiv:hep-ph/0105084].
  %%CITATION = HEP-PH 0105084;%%
 % D.~Chakraverty, K.~Huitu and A.~Kundu,
  %``Effects of universal extra dimensions on B0 - anti-B0 mixing,''
  %Phys.\ Lett.\ B {\bf 558}, 173 (2003).
  %[arXiv:hep-ph/0212047].
  %%CITATION = HEP-PH 0212047;%%

\bibitem{Appelquist:2000nn}
  T.~Appelquist, H.~C.~Cheng and B.~A.~Dobrescu,
  %``Bounds on universal extra dimensions,''
  Phys.\ Rev.\ D {\bf 64}, 035002 (2001).
%  [arXiv:hep-ph/0012100].
  %%CITATION = HEP-PH 0012100;%%

 \bibitem{Buras:2002ej}
  A.~J.~Buras, M.~Spranger and A.~Weiler,
  %``The impact of universal extra dimensions on the unitarity triangle and rare
  %K and B decays. ((U)),''
  Nucl.\ Phys.\ B {\bf 660}, 225 (2003).
%  [arXiv:hep-ph/0212143].
  %%CITATION = HEP-PH 0212143;%%

  \bibitem{Buras:2003mk}
  A.~J.~Buras {\it et al.},
  %``The impact of universal extra dimensions on B $\to$ X/s gamma, B $\to$ X/s
  %gluon, B $\to$ X/s mu+ mu-, K(L) $\to$ pi0 e+ e-, and epsilon'/epsilon,''
  Nucl.\ Phys.\ B {\bf 678}, 455 (2004).
%  [arXiv:hep-ph/0306158].
  %%CITATION = HEP-PH 0306158;%%

\bibitem{Colangelo:2006vm}
  P.~Colangelo, F.~De Fazio, R.~Ferrandes and T.~N.~Pham,
  %``Exclusive B $\to$ K(*) l+ l-, B $\to$ K(*) nu anti-nu and B $\to$ K* gamma
  %transitions in a scenario with a single universal extra dimension,''
  Phys.\ Rev.\ D {\bf 73}, 115006 (2006).
  %%CITATION = HEP-PH 0604029;%%

\bibitem{nnlo}
C.~Bobeth {\it et al.},
  %``Photonic penguins at two loops and m(t)-dependence of BR(B $\to$ X(s) l+
  %l-),''
  Nucl.\ Phys.\ B {\bf 574}, 291 (2000);
%  [arXiv:hep-ph/9910220];
  %%CITATION = HEP-PH 9910220;%%
  H.~H.~Asatrian {\it et al.},
  %``Two-loop virtual corrections to B $\to$ X/s l+ l- in the standard model,''
  Phys.\ Lett.\ B {\bf 507}, 162 (2001);
%  [arXiv:hep-ph/0103087];
  %%CITATION = HEP-PH 0103087;%%
  Phys.\ Rev.\ D {\bf 65}, 074004 (2002);
%  [arXiv:hep-ph/0109140];
  %%CITATION = HEP-PH 0109140;%%
  Phys.\ Rev.\ D {\bf 66}, 034009 (2002);
%  [arXiv:hep-ph/0204341];
  %%CITATION = HEP-PH 0204341;%%
  H.~M.~Asatrian {\it et al.},
  %``NNLL corrections to the angular distribution and to the forward-backward
  %asymmetries in b $\to$ X/s l+ l-,''
  Phys.\ Rev.\ D {\bf 66}, 094013 (2002);
%  [arXiv:hep-ph/0209006];
  %%CITATION = HEP-PH 0209006;%%
  A.~Ghinculov {\it et al.},
  %``Forward-backward asymmetry in B $\to$ X/s l+ l- at the NNLL level,''
  Nucl.\ Phys.\ B {\bf 648}, 254 (2003);
%  [arXiv:hep-ph/0208088].
  %%CITATION = HEP-PH 0208088;%%
  A.~Ghinculov {\it et al.},
 %``The rare decay B $\to$ X/s l+ l- to NNLL precision for arbitrary dilepton
 %invariant mass,''
 Nucl.\ Phys.\ B {\bf 685}, 351 (2004);
% [arXiv:hep-ph/0312128].
 %%CITATION = HEP-PH 0312128;%%
C.~Bobeth {\it et al.},
 %``Complete NNLO QCD analysis of anti-B $\to$ X/s l+ l- and higher order
 %electroweak effects,''
 JHEP {\bf 0404}, 071 (2004).
% [arXiv:hep-ph/0312090].
 %%CITATION = HEP-PH 0312090;%%

\bibitem{Colangelo:1995jv}
  P.~Colangelo {\it et al.},
  %``QCD Sum Rule Analysis of the Decays $B \to K \ell~+ \ell~-$ and $B \to K~*
  %\ell~+ \ell~-$,''
  Phys.\ Rev.\ D {\bf 53}, 3672 (1996)
  [Erratum-ibid.\ D {\bf 57}, 3186 (1998)].
%  [arXiv:hep-ph/9510403].
  %%CITATION = HEP-PH 9510403;%%

\bibitem{Ball:2004rg}
  P.~Ball and R.~Zwicky,
  %``B/(d,s) $\to$ rho, omega, K*, Phi decay form factors from light-cone sum
  %rules revisited,''
  Phys.\ Rev.\ D {\bf 71}, 014015 (2005);
  Phys.\ Rev.\ D {\bf 71}, 014029 (2005).
%  [arXiv:hep-ph/0412079].
  %%CITATION = HEP-PH 0412079;%%


\bibitem{Aubert:2004it}
  B.~Aubert {\it et al.}  [BaBar Collaboration],
  %``Measurement of the B $\to$ X/s l+ l- branching fraction with a sum over
  %exclusive modes,''
  Phys.\ Rev.\ Lett.\  {\bf 93}, 081802 (2004).
 % [arXiv:hep-ex/0404006].
  %%CITATION = HEP-EX 0404006;%%

\bibitem{Aubert:2005cf}
  B.~Aubert {\it et al.}  [BaBar Collaboration],
  %``Measurements of the rare decays B $\to$ K l+ l- and B $\to$ K* l+ l-,''
  arXiv:hep-ex/0507005.
  %%CITATION = HEP-EX 0507005;%%
\bibitem{Iwasaki:2005sy}
  M.~Iwasaki {\it et al.}  [Belle Collaboration],
  %``Improved measurement of the electroweak penguin process B $\to$ X/s l+
  %l-,''
  Phys.\ Rev.\ D {\bf 72}, 092005 (2005).
%  [arXiv:hep-ex/0503044].
  %%CITATION = HEP-EX 0503044;%%
\bibitem{Abe:2004ir}
  K.~Abe {\it et al.}  [Belle Collaboration],
  %``Measurement of the differential q**2 spectrum and forward-backward
  %asymmetry for B $\to$ K(*) l+ l-,''
  arXiv:hep-ex/0410006.
  %%CITATION = HEP-EX 0410006;%%

\bibitem{Burdman:1998mk}
  G.~Burdman,
  %``Short distance coefficients and the vanishing of the lepton asymmetry  in B
  %$\to$ V l+ l-,''
  Phys.\ Rev.\ D {\bf 57}, 4254 (1998).
%  [arXiv:hep-ph/9710550].

\bibitem{Abe:2005km}
 A.~Ishikawa {\it et al.},
  % ``Measurement of forward-backward asymmetry and Wilson coefficients in B -->
  %K* l+ l-,''
  Phys.\ Rev.\ Lett.\  {\bf 96}, 251801 (2006)
  [arXiv:hep-ex/0603018].
  %%CITATION = HEP-EX 0603018;%%

\bibitem{inami}
T.~Inami and C.~S.~Lim,
  %``Effects Of Superheavy Quarks And Leptons In Low-Energy Weak Processes K(L)
  %$\to$ Mu Anti-Mu, K+ $\to$ Pi+ Neutrino Anti-Neutrino And K0 <---> Anti-K0,''
  Prog.\ Theor.\ Phys.\  {\bf 65}, 297 (1981)
  [Erratum-ibid.\  {\bf 65}, 1772 (1981)].
  %%CITATION = PTPKA,65,297;%%

\bibitem{buchalla}
G.~Buchalla and A.~J.~Buras,
  %``QCD corrections to rare K and B decays for arbitrary top quark mass,''
  Nucl.\ Phys.\ B {\bf 400}, 225 (1993).
  %%CITATION = NUPHA,B400,225;%%

\bibitem{buchalla1}
G.~Buchalla {\it et al.},
  %``Weak Decays Beyond Leading Logarithms,''
  Rev.\ Mod.\ Phys.\  {\bf 68}, 1125 (1996).
%  [arXiv:hep-ph/9512380].
  %%CITATION = HEP-PH 9512380;%%

%\cite{Buchalla:1998ba}
\bibitem{Buchalla:1998ba}
  G.~Buchalla and A.~J.~Buras,
  %``The rare decays K $\to$ pi nu anti-nu, B $\to$ X nu anti-nu and  B $\to$ l+
  %l-: An update,''
  Nucl.\ Phys.\ B {\bf 548}, 309 (1999).
  %[arXiv:hep-ph/9901288].
  %%CITATION = HEP-PH 9901288;%%

  \bibitem{Colangelo:1996ay}
  P.~Colangelo {\it et al.},
  %``Rare B $\to$ K(*) nu anti-nu decays at B factories,''
  Phys.\ Lett.\ B {\bf 395}, 339 (1997).
%  [arXiv:hep-ph/9610297].
  %%CITATION = HEP-PH 9610297;%%

\bibitem{Buchalla:2000sk}
  G.~Buchalla, G.~Hiller and G.~Isidori,
  %``Phenomenology of non-standard Z couplings in exclusive semileptonic  b
  %$\to$ s transitions,''
  Phys.\ Rev.\ D {\bf 63}, 014015 (2001).
%  [arXiv:hep-ph/0006136].
  %%CITATION = HEP-PH 0006136;%%
\bibitem{Abe:2005bq}
  K.~Abe {\it et al.}  [Belle Collaboration],
  %``Search for B $\to$ tau nu and B $\to$ K nu anti-nu decays with a fully
  %reconstructed B at belle,''
  arXiv:hep-ex/0507034.
  %%CITATION = HEP-EX 0507034;%%

\bibitem{Aubert:2004ws}
  B.~Aubert {\it et al.}  [BaBar Collaboration],
  %``A search for the decay B+ $\to$ K+ nu anti-nu,''
  Phys.\ Rev.\ Lett.\  {\bf 94}, 101801 (2005).
%  [arXiv:hep-ex/0411061].
  %%CITATION = HEP-EX 0411061;%%

  \bibitem{Nakao:2004th}
  M.~Nakao {\it et al.}  [Belle Collaboration],
  %``Measurement of the B $\to$ K* gamma branching fractions and asymmetries,''
  Phys.\ Rev.\ D {\bf 69}, 112001 (2004).
%  [arXiv:hep-ex/0402042].
 %%CITATION = HEP-EX 0402042;%%

\bibitem{Aubert:2004te}
  B.~Aubert {\it et al.}  [BaBar Collaboration],
  %``Measurement of branching fractions, and CP and isospin asymmetries, for  B
  %$\to$ K* gamma,''
  Phys.\ Rev.\ D {\bf 70}, 112006 (2004).
%  [arXiv:hep-ex/0407003].
 %%CITATION = HEP-EX 0407003;%%
%%%%%%%%%%%%%%%%%%%%%%%%%%%%%%%%%%%%%%%%%%%%%%%%%%%%%%%%%%%%%%%%%%%%%%%%%%%%%%%%%%%%%%%%%%



\end{thebibliography}
\end{document}